\documentclass[aip,amsmath,amssymb,superscriptaddress,reprint]{revtex4-1}

\usepackage{graphicx}
\usepackage{bm}
\usepackage{multirow}
\usepackage{verbatim}
\usepackage{longtable}
\usepackage{color}
\usepackage{hyperref}
\usepackage[usenames,dvipsnames]{xcolor}
\usepackage{csquotes}

\begin{document}

\title{Electroabsorption in gated GaAs nanophotonic waveguides}
\author{Ying Wang}
\affiliation{Center for Hybrid Quantum Networks (Hy-Q), Niels Bohr Institute, University of Copenhagen, Blegdamsvej 17, DK-2100 Copenhagen, Denmark}
\author{Ravitej Uppu}
\affiliation{Center for Hybrid Quantum Networks (Hy-Q), Niels Bohr Institute, University of Copenhagen, Blegdamsvej 17, DK-2100 Copenhagen, Denmark}
\author{Xiaoyan Zhou}
\affiliation{Center for Hybrid Quantum Networks (Hy-Q), Niels Bohr Institute, University of Copenhagen, Blegdamsvej 17, DK-2100 Copenhagen, Denmark}
\author{Camille Papon}
\affiliation{Center for Hybrid Quantum Networks (Hy-Q), Niels Bohr Institute, University of Copenhagen, Blegdamsvej 17, DK-2100 Copenhagen, Denmark}
\author{Sven Scholz}
\affiliation{Lehrstuhl f{\"u}r Angewandte Festk{\"o}rperphysik, Ruhr-Universit{\"a}t Bochum, Universit{\"a}tsstrasse 150, D-44780 Bochum, Germany}
\author{Andreas D.~Wieck}
\affiliation{Lehrstuhl f{\"u}r Angewandte Festk{\"o}rperphysik, Ruhr-Universit{\"a}t Bochum, Universit{\"a}tsstrasse 150, D-44780 Bochum, Germany}
\author{Arne Ludwig}
\affiliation{Lehrstuhl f{\"u}r Angewandte Festk{\"o}rperphysik, Ruhr-Universit{\"a}t Bochum, Universit{\"a}tsstrasse 150, D-44780 Bochum, Germany}
\author{Peter Lodahl}
\affiliation{Center for Hybrid Quantum Networks (Hy-Q), Niels Bohr Institute, University of Copenhagen, Blegdamsvej 17, DK-2100 Copenhagen, Denmark}
\author{Leonardo Midolo}
\email[Author to whom correspondence should be addressed:]{midolo@nbi.ku.dk}
\affiliation{Center for Hybrid Quantum Networks (Hy-Q), Niels Bohr Institute, University of Copenhagen, Blegdamsvej 17, DK-2100 Copenhagen, Denmark}

\begin{abstract}
We report on the analysis of electroabsorption in thin GaAs/Al$_{0.3}$Ga$_{0.7}$As nanophotonic waveguides with an embedded $p$-$i$-$n$ junction. By measuring the transmission through waveguides of different lengths, we derive the propagation loss as a function of electric field, wavelength, and temperature. The results are in good agreement with the Franz-Keldysh model of electroabsorption extending over 200 meV below the GaAs bandgap, i.e. in the 910--970 nm wavelength range. We find a pronounced residual absorption in forward bias, which we attribute to Fermi-level pinning at the waveguide surface, producing over 20 dB/mm loss at room temperature. These results are essential for understanding the origin of loss in nanophotonic devices operating in the emission range of self-assembled InAs semiconductor quantum dots, towards the realization of scalable quantum photonic integrated circuits.
\end{abstract}
\maketitle 

An ambitious goal in photonic quantum technologies is to scale up photonic integrated circuits for generating, routing, and detecting single photons efficiently \cite{obrien_photonic_2009}. One of the fundamental challenges is to keep the overall device and circuit loss as low as possible, as the fundamental no-cloning theorem of quantum mechanics forbids amplification of the single-photon signal. 
Gallium arsenide (GaAs) membranes with embedded self-assembled quantum dots (QDs) constitute a mature platform for integrating deterministic single-photon emitters with planar photonic integrated circuits\cite{hepp2019semiconductor,lodahl2017quantum}, with excellent single-photon source performance. \cite{uppu2020chip,uppu2020scalable} 
Furthermore, the integration of doped layers in the epitaxial heterostructure, enables building thin $p$-$i$-$n$ diode junctions to control the state and charge occupation of the QD to reduce electrical noise\cite{wang2016near,pedersen2020near,ollivier2020reproducibility} and to tune the emitter wavelength via the quantum-confined Stark effect.\cite{laucht2009electrical} At the same time, however, the doped layers introduce additional loss due to free-carrier absorption and the Franz-Keldysh effect (FKE).\cite{rosencher2002optoelectronics} The latter originates from the strong built-in electric fields in such thin junctions, which distort the electron and hole wave-functions in GaAs and enable optical absorption at energies below the bandgap. 

\begin{figure}[h]
\centering
\includegraphics[width=8.4cm]{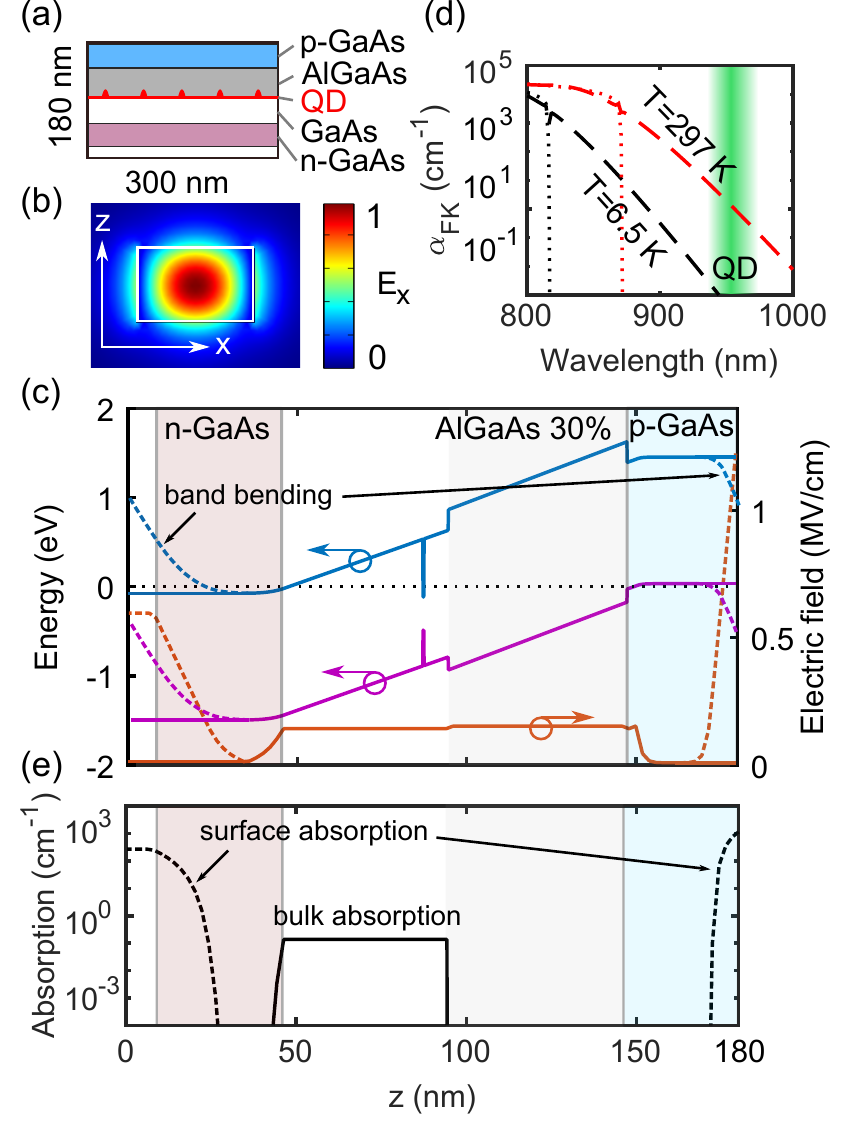}
\caption{\label{fig:f1} Electroabsorption in GaAs nanophotonic waveguides. (a) Layout of the p-i-n heterostructure used to fabricate the device.
(b) Finite element method simulation of the transverse electric field mode profile. (c) Band diagram at equilibrium (without external bias). The blue (purple) solid line represents the conduction (valence) band and the dotted line is the Fermi level. The built-in electric field is plotted on the right axis (red solid line). Dashed lines show the case of band bending due to Fermi-level pinning. (d) Electroabsorption as a function of wavelength for cryogenic and room temperature. The dashed (dotted) line indicates the absorption for an electric field $F=100 $ kV/cm ($F=0$ kV/cm). The green area indicates the typical emission range of In(Ga)As QDs.  (e) Electroabsorption profile due to Franz-Keldysh effect calculated from the electric field in (c) at a wavelength of 930 nm and $T=6.5$ K. A large absorption is expected at the surface, due to Fermi-level pinning.}
\end{figure}
In this work, we report on a detailed analysis of the various attenuation mechanisms in doped GaAs waveguides. We analyze the impact of electroabsorption and surface states on the waveguide loss. From the observed results, we conclude that a significant contribution to electroabsorption originates from strong band-bending at the surface of the waveguide, which is caused by mid-gap surface states. Previous works suggested that loss in doped waveguides can arise from unpassivated surfaces,\cite{najer2019gated,midolo2017electro} but no quantitative analysis has been reported. By comparing measurements with and without doped layers and at different temperatures, we can identify the origin of loss and devise a strategy for designing and fabricating the next generation of quantum photonic integrated devices with quantum emitters.

The devices used in this work are single-mode rectangular waveguides fabricated on a 180-nm-thick GaAs/Al$_{0.3}$Ga$_{0.7}$As membrane with a width of 300 nm. The layout of the structure and the profile of the fundamental transverse electric (TE) mode, $E_x(x,z)$, are shown in Fig. \ref{fig:f1}(a) and \ref{fig:f1}(b), respectively.
The heterostructure contains a $p$-$i$-$n$ diode with a 100-nm-thick undoped region, with a layer of self-assembled InAs QDs located in the middle. The Al$_{0.3}$Ga$_{0.7}$As layer above the QDs serves as barrier to reduce the tunneling rate of holes out of the QDs, thereby extending the Stark tuning range of the emitter. \cite{bennett2010giant} Further details on the heterostructure used in this work are given in ref.\cite{uppu2020chip}.
Figure \ref{fig:f1}(c) shows the diode band diagram at $T=6.5$ K, calculated with a one-dimensional Poisson equation solver, \cite{tan1990self} when no external voltage is applied to the diode.
The solid lines show the conduction and valence band in the case of perfect ohmic contacts to the $p$-type and $n$-type layers. In a more realistic scenario, the waveguide surfaces have been exposed to air and have oxidized, producing a large amount of defects whose energy level is located within the GaAs bandgap. These mid-gap states are populated causing a net increase of surface charge density. The presence of surface defects is known to ``pin'' the Fermi level inside the gap, causing band bending in the proximity of the surface.\cite{colleoni2015fermi} To model the effect of the Fermi-level pinning, we apply a Schottky-type boundary condition with a barrier height of $1$ eV on both sides of the diode. The Schottky barrier height is estimated assuming a density of surface states $\sim 10^{13}$ cm$^{-2}$,\cite{offsey1986unpinned} which pins the Fermi level $\sim 0.5$ eV above the valence band maximum, as reported in previous experiments.\cite{spicer1979new} The bending of conduction and valence bands is shown in Fig.~\ref{fig:f1}(c) as dashed lines.
The corresponding electric field (with and without band bending) is plotted in Fig. \ref{fig:f1}(c) (right axis). Fields exceeding 1 MV/cm appear at the top surface due to Fermi-level pinning. Such large fields are expected to contribute significantly to the absorption due to the FKE, even several hundreds of meV inside the gap, where excitonic states of QDs are located.

To model the Franz-Keldysh absorption $\alpha_\textrm{FK}(\lambda,F)$ as a function of wavelength $\lambda$ and electric field $F$, we use the theory by Callaway,\cite{callaway1964optical} with the phenomenological scaling factors reported in ref. \cite{stillman1976electroabsorption}. 
These models and experiments cover the wavelengths (910--970 nm) and the electric field ranges of our experiment, but have been only experimentally validated at room temperature and in bulk GaAs. We assume that a change of temperature only affects the bandgap of GaAs, shifting it from $E_\textrm{g} = 1.44$ eV at room temperature to $E_\textrm{g} = 1.52$ eV at $T=6.5$ K, resulting in a lower absorption at cryogenic conditions. 
The calculated absorption for the two temperatures used in this work, together with the inhomogeneous distribution of QD emission, is shown in Fig. \ref{fig:f1}(d). The dotted lines show the typical zero-field absorption of direct-gap undoped semiconductors, i.e. when no absorption is expected at energies below the bandgap. The dashed lines show instead the absorption when $F=100$ kV/cm, i.e. when the FKE becomes significant in the QD emission region. In the model, the absorption due to QDs themselves has been neglected since the QD density $\simeq 10 \mu$m$^{-2}$ and their absorption cross-section are too small to be observed when compared to all other loss mechanisms. Moreover, the quantum-confined Stark effect from QDs can also be neglected in our analysis, as the emitters' inhomogeneous broadening ($>30$ meV) is much larger than the expected Stark tuning (in the order of 1 meV/V).

From the electric field profile $F(z)$ calculated in Fig. \ref{fig:f1}(c) and the FKE model $\alpha_\textrm{FK}\left(\lambda,F(z)\right)$, a profile of the electroabsorption as a function of the position $z$ in the waveguide can be derived and it is plotted in Fig. \ref{fig:f1}(e). Here, we only plot the absorption due to the built-in electric field ($V_\textrm{BI}\simeq1.52$ V) and surface fields due to Fermi-level pinning, while we neglect the electroabsorption in the Al$_{0.3}$Ga$_{0.7}$As region, as the bandgap is further blue-shifted to 1.77 eV.  
We further separate the contribution to the total absorption in two spatially distinct regions: 1) the undoped region inside the $p$-$i$-$n$ junction, where the absorption depends on the externally-applied bias (solid line in Fig.~\ref{fig:f1}(e)) and 2) the surface regions, where the absorption is only due to Fermi-level pinning and does not depend significantly on the external field (dashed lines). 
We model the total loss in the waveguide as
\begin{equation}
\label{eq:fitmodel}
\alpha(\lambda,F)=\alpha_\textrm{R}(\lambda)+\Gamma_\textrm{b}\alpha_\textrm{FK}(\lambda,F),
\end{equation}
where $\alpha_\textrm{R}(\lambda)$ is the residual absorption (i.e. without the voltage-dependent FKE) and $\Gamma_\textrm{b}$ is the mode confinement factor\cite{visser1997confinement} for the undoped GaAs region, given by the overlap integral
\begin{equation}
\Gamma_\textrm{b} = \frac{c}{2n}\frac{\int_\textrm{b} \varepsilon(x,z)\left|\vec{E}(x,z)\right|^2 dxdz}{\int  \langle S_y(x,z)\rangle dxdz}.
\label{eq:overlap}
\end{equation}
Here, $\vec{E}(x,z)$ and $\langle S_y(x,z)\rangle$ are the simulated two-dimensional profiles of the optical mode electric field (cf. Fig.\ref{fig:f1}b) and of the average Poynting vector  component in the propagation direction of the waveguide, respectively, while $\varepsilon(x,z)$ is the spatial distribution of the permittivity. The subscript b denotes the undoped (bulk) GaAs layer in the waveguide.
The electric field in the undoped region is calculated using $F=(V_\textrm{BI}-V)/d$, where $d$ is the total length of the undoped and depletion layer and $V_\textrm{BI}$ is the built-in voltage of the diode junction.

\begin{figure}[htb!]
\centering
\includegraphics[width=7.4cm]{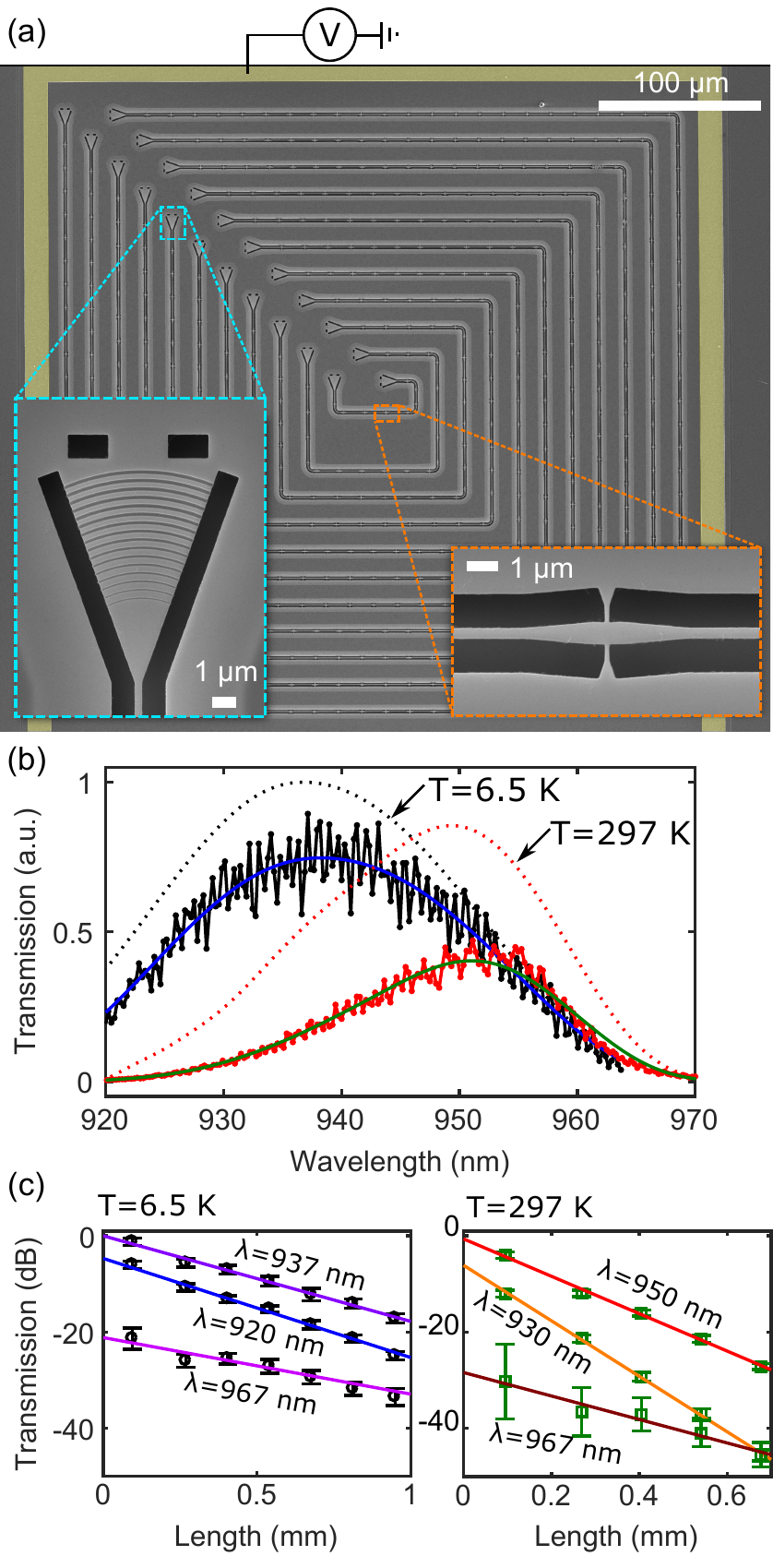}
\caption{\label{fig:f2} Measurement of waveguide attenuation. (a) Scanning electron microscope (SEM) image of the suspended concentric waveguides. Details of the waveguide-supporting tethers and focusing gratings are shown in the insets. Electrical contacts (highlighted in yellow) are used to apply an electric field to the entire structure.
(b) Transmission spectra at room and low temperature (red and black dots connected with solid lines, respectively) without applied bias ($L=95$ $\mu$m). The solid lines show smoothed data while the dotted lines show the transmission spectrum of the gratings $T_\textrm{G}$. All plots are normalized to the peak transmission of the grating at low temperature.
(c) Transmission (in dB) normalized to the peak transmission of the grating as a function of waveguide length at low (left) and room (right) temperature. The solid lines are linear fits where the slope represents the loss per unit length.}
\end{figure}

To measure the loss per unit length in nanophotonic waveguides, a series of concentric waveguides have been fabricated on the sample following the procedure outlined in a previous work.\cite{midolo2015soft} The gratings are written with electron-beam lithography and etched via reactive ion etching in the same run, to ensure high reproducibility of their transmission spectrum within a 10\% error. Figure \ref{fig:f2}(a) shows a top-view scanning-electron microscope (SEM) image of the fabricated sample. The p-contact electrode is highlighted in yellow and it is used to apply a voltage to the $p$-$i$-$n$ junction in the GaAs membranes. A pair of gratings\cite{zhou2018high} are used to couple light in and out of the waveguides. The gratings are placed at a fixed distance from each other, to avoid re-aligning the position and angle of excitation and collection beams, and cross-polarized to avoid collecting any scattered light from the input ports. This method allows factoring out the loss due to the gratings and the three 90-degree bends. Several tethers are used to keep the waveguide suspended in air, as shown in the inset of Fig. \ref{fig:f2}(a). Varying the number of tethers per unit length does not affect the transmission in our experiments,\cite{papon2019nanomechanical} therefore we assume all tethers to be lossless.
The sample is mounted in a closed-cycle cryostat and transmission measurements are performed using a supercontinuum laser source. Figure \ref{fig:f2}(b) shows the transmission spectra of the shortest waveguide $L=95$ $\mu$m, at cryogenic and room temperature, without any bias applied to the diode. A visible quenching of the transmission is observed at room temperature, due to the built-in electroabsorption. Due to the thermo-optic effect, the grating transmission peak red-shifts by roughly 20 nm at room temperature.
The small fringes, likely caused by the small reflectivity of the grating coupler, are smoothed out to extract a consistent response function at each wavelength (see solid lines in Fig. \ref{fig:f2}(b)). The fringe amplitude is used as an error bar for the transmission data. We note that the error bar typically exceeds the grating-to-grating variations due to fabrication and alignment errors. Figure \ref{fig:f2}(c) shows the transmission, in logarithmic scale, as a function of the waveguide length for both room and low temperatures. The attenuation coefficient $\alpha$ is obtained by a linear fit of the logarithm of the transmission according to
\begin{equation}
\log(T(L)) = 3 \log(T_\textrm{B}) + 2 \log(T_\textrm{G}) - \alpha L,
\label{eq:attenuate}
\end{equation}
where $T_\textrm{G}$ is the transmission of a single grating, derived from the intercept of the transmission curve with the ordinate, i.e. at zero length and $T_\textrm{B}$ is the transmission of a single 90-degree bend, which, from numerical simulations, accounts for $\sim 7 \mu$m waveguide length and $<0.005$ dB loss in the wavelength range investigated here. A loss per unit length of (17$\pm$ 1) dB/mm and (39 $\pm$ 3) dB/mm is obtained at cryogenic and room temperature, respectively for the wavelengths of 937 nm and 950 nm, i.e. at the transmission peak of the gratings. Additionally, two fits are shown for wavelengths away from the central peak, where transmission is typically lower. Yet, the transmission in dB shows a linear dependence on the waveguide length within error bars, allowing us to extract a loss per unit length over a broader wavelength range. The grating transmission $T_\textrm{G}$, obtained by the extrapolation described above, is shown as a function of wavelength in Fig. \ref{fig:f2}(b) as dotted lines. The grating transmission is compatible with results previously reported for undoped waveguides. \cite{zhou2018high} Also, the wavelength-dependence of the FKE is already visible by comparing the grating response and the measured data in Fig. \ref{fig:f2}(c): a more pronounced absorption is observed at shorter wavelengths than at longer wavelengths, as expected from theory (cf. Fig. \ref{fig:f1}(d)).

\begin{figure}[h]
\centering
\includegraphics[width=7.4cm]{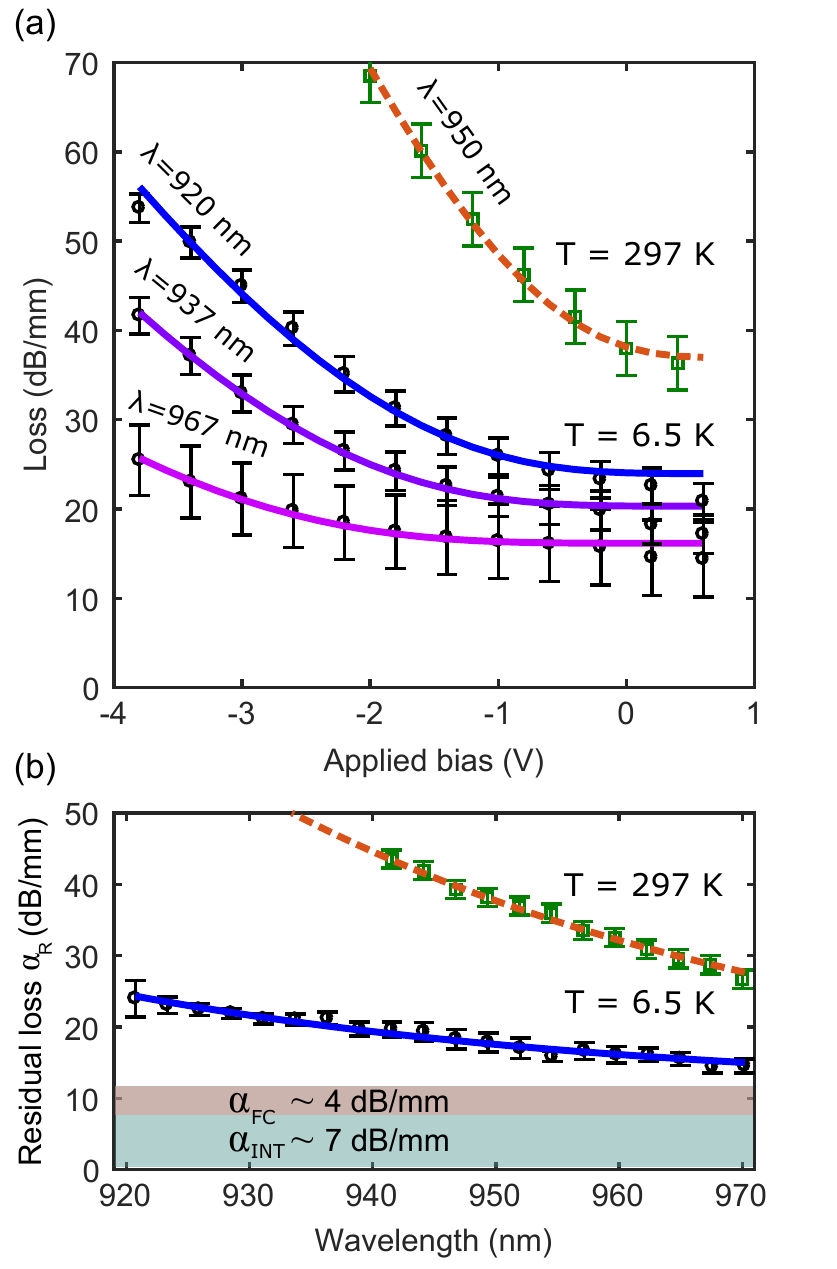}
\caption{\label{fig:f3} Voltage and wavelength dependance of waveguide attenuation. (a) Waveguide loss as a function of the applied bias. The bottom three curves (solid lines) are fit to data collected at low temperature (T=6.5 K) and at different wavelengths (black circles). The loss at room temperature (green squares) and the fit (red dashed line) are also shown for comparison. (b) Residual loss extrapolated from the fits in (a) as a function of the wavelength. The low-temperature (black circles) and room-temperature (green squares) data are fitted with an electroabsorption model with a fixed surface electric field of 0.65 MV/cm. The remaining loss contribution, indicated by the shaded areas is due to free-carrier (FC) absorption and intrinsic losses with negligible dependence on wavelength and temperature.}
\end{figure}

To identify the various contributions to loss in doped waveguides, the applied bias is varied from $V=+0.6$ V (forward bias) to $V=-4$ V (reverse bias) in steps of $0.4$ V and the measurement and fitting procedure described in Fig. \ref{fig:f2} is repeated at all voltages. Figure \ref{fig:f3}(a) shows the dependence of the waveguide loss $\alpha$ as a function of the applied bias for three different wavelengths at $T=6.5$ K and for the peak wavelength $\lambda=950$ nm at room temperature. Error bars are given in logarithmic scale by propagation of uncertainties after fitting. 
A non-linear least-square fitting procedure is used to fit the data with the model of equation \ref{eq:fitmodel}, where the fitting parameters are the confinement factor of the undoped region $\Gamma_\textrm{b}$, and the residual loss.
The residual loss $\alpha_\textrm{R}$ is plotted in Fig. \ref{fig:f3}(b) as a function of wavelength and temperature. This is the loss which remains after compensating the built-in voltage in the waveguide. The pronounced wavelength-dependence of such residual loss suggests that an additional FKE is present, most likely due to the surface fields caused by Fermi-level pinning. We note that the loss per unit length extracted at wavelengths away from the grating peak would generally be more sensitive to variations between different samples. Nonetheless, all data analysis was carried out in a parameter range where such sample variations were minor compared to the extracted propagation loss. 

To confirm this, the residual loss is fitted as a function of wavelength using the FKE model:
\begin{equation}
\label{eq:fitmodel2}
\alpha_\textrm{R}(\lambda)=\alpha_0+\Gamma_\textrm{s}\alpha_\textrm{FK}(\lambda,F_\textrm{s}),
\end{equation}
where we use a fixed field of $F_\textrm{s}=0.65$ MV/cm obtained from the simulated average surface field (cf. dashed line of Fig.\ref{fig:f1}(c)). From the fit, the confinement factor of the surface region $\Gamma_\textrm{s}$ and the background loss $\alpha_0$ are extracted. The solid lines show the best fit to the data. The results of the two fits of Fig. \ref{fig:f3} are summarized in table \ref{tab:t1}, and compared to theoretical values obtained from finite-element calculations.
\begin{table}[hbt]
\centering
\begin{tabular}{| c | c | c | c |}
		\hline
      & $T=297$ K & $T=6.5$ K & Theory\\
    \hline

     $\Gamma_b$ & $(47 \pm 7)$\% & $(51 \pm 7)$\%  & 52 \% \\ 
     $\Gamma_s$ & $(6.6 \pm 0.5)$\% & $(6.3 \pm 0.5)$\% & 4 \% \\
     $\alpha_0$ & $(10.4 \pm 0.9)$ dB/mm & $(10.9 \pm 1)$ dB/mm & - \\
		\hline
		$\alpha_{FC}$ & \multicolumn{3}{c|}{$\sim 4$ dB/mm (estimated)\cite{casey1975concentration}} \\
		$\alpha_{INT}$ & \multicolumn{3}{c|}{$\sim 7$ dB/mm (measured)\cite{papon2019nanomechanical}} \\
		\hline
\end{tabular}
\caption{\label{tab:t1} Results of the fits of Fig.~\ref{fig:f3}.}
\end{table}

Both low-temperature and room-temperature fits result in a constant background loss of $\alpha_0 = (11 \pm 2)$ dB/mm. Such loss can be attributed to sum of two main contributions, namely free-carrier absorption ($\alpha_\textrm{FC}$) and intrinsic loss ($\alpha_\textrm{INT}$). Free-carrier absorption is estimated using the data for $p$-doped and $n$-doped layers given in ref \cite{casey1975concentration} and scaled by the confinement factors of equation \ref{eq:overlap}, resulting in $\alpha_\textrm{FC}\sim 4$ dB/mm. Intrinsic loss of waveguide is measured on a separate experiment that uses undoped wafers but identical waveguide geometry. The results have been previously published in ref.\cite{papon2019nanomechanical} and resulted in $\alpha_\textrm{INT}\sim 7$ dB/mm. The main source of intrinsic loss is suspected to be caused by scattering due to sidewall roughness arising in the GaAs etching process. The root-mean square (RMS) roughness in our waveguides is estimated to be in the $\sim 3$--$4$ nm range from SEM analysis. According to waveguide scattering theory such as the Payne-Lacey model,\cite{payne1994theoretical} roughness causes unwanted coupling between the fundamental guided mode and radiation modes, which is stronger for high-index contrast waveguides. A full three-dimensional finite-element simulation of transmission over rough waveguides provides a loss in the 5--8 dB/mm range, consistent with our experimental findings. Another potential loss mechanisms is given by the aforementioned mid-gap states which, even in the absence of doped layers, can cause absorption below the bandgap energy. However, further experiments are required to reduce the scattering loss and observe such loss mechanisms.

In conclusion, we have reported the analysis of electroabsorption in gated GaAs nanophotonic waveguides as a function of voltage, wavelength, and temperature. While the fundamental building blocks and devices for quantum applications can be realized over small ($<100$ $\mu$m) dimensions, where the losses are negligible, scaling up photonic integrated circuits in doped GaAs requires a deep understanding of the origin of loss. To this end, it is crucial to devise a method to remove the sources of absorption while still operating at $\lambda \sim 930$ nm, i.e. the central wavelength of high-quality QDs. The $p$-doped layer is the largest source of free-carrier absorption and it is only required on the emitter region. By selectively etching away the $p$-layer from the surface, the free-carrier absorption can be reduced to $\alpha_\textrm{FC}\sim 0.5$ dB/mm (i.e. the $n$-layer contribution) and the FKE due to built-in fields can be suppressed. Surface fields, however, require strategies for unpinning the Fermi level. Several works have shown that oxide removal followed by surface passivation, greatly enhances the quality factor of microdisks, photonic crystal cavities and vertical microcavities.\cite{guha2017surface,najer2019gated,kuruma2020surface} By removing mid-gap states, it is expected that the surface Fermi level unpins, thereby removing any source of electroabsorption due to band-bending. 
Finally, improving the fabrication process using intermediate hard masks \cite{gonzalez2014fabrication} or resist reflow \cite{benevides2020ar} is also expected to reduce the waveguide RMS roughness to $< 1$ nm, which can potentially reduce the intrinsic waveguide loss down to $\alpha_\textrm{INT}< 1$ dB/mm. Achieving such a target would enable integrating several hundreds of devices such as single-photon sources, filters, and switches, while taking advantage of the small size of such devices in the GaAs platform. For example, recently reported nanomechanical switches and filters,\cite{papon2019nanomechanical,zhou2019chip} have footprints of $\sim 25$ $\mu$m, which would allow implementing large ($>40$-mode) unitary gates with $\sim 1$ mm optical depth, required for photonic quantum information processing.

\section*{Data availability statement}
The data that support the findings of this study are available from the corresponding author upon reasonable request.

\begin{acknowledgments}
We acknowledge Matthias L\"{o}bl, Alisa Javadi, and Richard J. Warburton for useful discussions. We gratefully acknowledge financial support from the Chinese Scholarship Council (CSC), Danmarks Grundforskningsfond (DNRF 139, Hy-Q Center for Hybrid Quantum Networks), H2020 European Research Council (ERC) (SCALE), Styrelsen for Forskning og Innovation (FI) (5072-00016B QUANTECH), Innovation Fund Denmark (9090-00031B, FIRE-Q), Bundesministerium f\"{u}r Bildung und Forschung (BMBF) (16KIS0867, Q.Link.X), Deutsche Forschungsgemeinschaft (DFG) (TRR 160).
\end{acknowledgments}

\nocite{*}

\end{document}